\documentstyle[preprint,aps]{revtex}
\begin{document}
\draft
\title{Orientation of particle attachment and local isotropy
in diffusion limited aggregation (DLA)}
\author{
Chi-Hang Lam}
\address{Department of Applied Physics, Hong Kong Polytechnic, Hung
Hom, Hong Kong.}
\maketitle
\begin{abstract}
We simulate 50 off-lattice DLA clusters, one million particles each.
The probability distribution of the angle of attachment of arriving
particles with respect to the local radial direction is obtained
numerically. For increasing cluster size, $N$, the distribution
crosses over extremely accurately to a cosine, whose amplitude
decreases towards zero as a power-law in $N$.  From this viewpoint,
asymptotically large DLA clusters are locally $isotropic$.  This
contradicts previous conclusions drawn from density-density
correlation measurements [P. Meakin, and T. Viscek, Phys. Rev. A {\bf
32}, 685 (1985)]. We present an intuitive phenomenological model
random process for our numerical findings.
\end{abstract}


\narrowtext

\section{Introduction}

The diffusion limited aggregation (DLA) \cite{Witten} is a fractal
growth model exhibiting great complexity \cite{Viscekbook}. The properties
of asymptotically large clusters are of fundamental importance. Some
of these properties are independent of many details of the aggregation
rules and are expected to be shared by large clusters grown in various
experimental or natural situations.  Efficient algorithms and improved
computational facilities has enabled the generation of very large
off-lattice clusters with more than 100 million particles
\cite{Kaufman}.  However, many asymptotic properties of DLA
are still unclear. There have been many debates on basic issues
including the scaling behavior of large clusters and the multi-fractal
properties of the growth probability measure
\cite{Viscekbook}.

This work concentrates on one of the most fundamental geometrical
aspects, namely the local isotropy, of large off-lattice DLA. Meakin
and Viscek \cite{Meakin_corr} found that the two points
density-density correlation is anisotropic. The tangential correlation
decays algebraically with exponent $\alpha_\perp \simeq 0.41$, which
is different from the radial exponent $\alpha_\parallel \simeq 0.29$.
As a result, the geometry, in particular the orientation, of a segment
of the cluster is related to its position inside the cluster. This
behavior does not hold for some deterministic fractal curves
containing spirals such as the Koch curve \cite{Mandelbrotbook} in the
limit of small segments.  Asymptotically large DLA would be locally
isotropic only if the two very different exponents converge to the
same value. Such a trend was not identified \cite{Meakin_corr} and the
anisotropy has been assumed to persist asymptotically
\cite{Meakin_corr,Mandelbrot}. Here, we apply a substantially more accurate
method and obtain the first unambiguously evidence of a systematic
decrease in the local anisotropy of DLA for increasing cluster size.
The trend of the decrease strongly indicates that asymptotically large
clusters are locally $isotropic$, contradictory to previous
conclusions
\cite{Meakin_corr}.

Specifically, we investigate the orientation of particle attachment.
For each new particle in the cluster, let $\vec{R}'$ and $\vec{R}$ be
the position vectors relative to the center the cluster of
respectively the new particle and its parent, where the parent is the
particle in the aggregate upon which the new attachment is made. We
define the center of the cluster to be at the seed.  The vector
$\vec{R}$ characterizes the local radial direction of this sticking
event and $\vec{r}=\vec{R}'-\vec{R}$ gives the sticking direction.
The angle of attachment $\theta$ ($-\pi < \theta \le \pi$) is defined
as the angle measured counter-clockwise from $\vec{R}$ to $\vec{r}$.
Attachment with with $\mid \theta
\mid ~<\pi/2$ can be described as forward.
We focus on the probability distribution $P(\theta,N)$ of $\theta$,
which, in general, depends on the number of particles, $N$, in the
cluster.  If DLA were compact with nearly circular boundary, the above
definition implies that $\theta$ is always close to zero and a
backward attachment is geometrically impossible. The distribution
$P(\theta,N)$ should peak at $\theta=0$.  Although, DLA is far from
being compact, at least for small clusters, forward attachment is
favored for similar geometrical reason.

\section{Simulations and results}

We compute the distribution $P(\theta,N)$ for $N= 150 \times 2^k$,
where k=0 to 12 corresponding to N=150 to 614400. For each $k$, we
histogram the values of the sticking angle $\theta$ for the
$(100\times 2^k+1$)-th to the $(200\times 2^k)$-th particles. Proper
normalization gives $P(\theta,N)$.  The symmetry
$P(\theta,N)=P(-\theta,N)$ is used.  We averaged the results over 50
clusters.  The statistical error is estimated from the sample to
sample fluctuations.

Figure \ref{F_Prob} shows $P(\theta,N)$ for 5 values of $N$. We used
16 bins to histogram. The error bars are smaller than the symbols
except for $N=150$. The errors are as small as about $0.2\%$ at
$N=614400$, where our data is most accurate. For small $N$, forward
sticking dominates so that $P(\theta,N)$ peaks at $\theta=0$ as expected.
Backward stickings are rare.  The peak broadens as $N$ increases.  At
$N=614400$, the ratio $P(\pm\pi,N)/P(0,N)$ between perfectly backward
and forward attachments becomes about 0.31. Figure
\ref{F_Prob} show least square fits to:
\begin{equation}
\label{P}
P(\theta,N)= 1/2\pi+ a_1(N)~ cos(\theta)
\end{equation}
for the 3 larger values of $N$, where $1/2\pi$ ensures normalization.
At $N=614400$, the quality of the fit amazing. There is apparently no
systematic trend of deviation, as the data points scatter around the
fitted curve by amounts comparable to the statistical errors.  The fit
is still good at $N=76800$, but at $N=9600$, there are noticeable
systematic deviations. The quality of the fit is restored when we
consider one more term in the cosine expansion:
\begin{equation}
\label{Pexpan}
P(\theta,N)= \frac{1}{2\pi} + \sum_{n=1}^{\infty}a_n(N)~ cos(n\theta)
\end{equation}
Figure \ref{F_Prob} shows the two parameters fits for $N=150$ and 1200.
More generally, we fitted our complete data set of
$P(\theta,N)$ with the first 4 cosine terms.
Figure \ref{F_amp}(a) plots the values of the amplitudes $a_n(N)$ in
semi-log. We used histograms of 64 bins and the result agrees with
those with 16 bins. We also computed $a_n(N)$ by Fourier transform and
obtained again the same results. Figure
\ref{F_amp}(b) and (c) shows respectively $a_1(N)$ and $a_2(N)$ in
log-log plots. The linearity at $N \agt 300$ implies:
\begin{equation}
\label{a_n}
a_n(N)\simeq A_n N^{-\gamma_n}
\end{equation}
for $n=1$ and 2, where $\gamma_1=0.0997(3), A_1=0.309(2),
\gamma_2=0.67(3)$ and $A_2=1.2(3)$. The bracketed values
are the fitting errors.  For $n\ge 3$, the measured $a_n(N)$ is not
precise enough for a test of the above algebraic decay.

We also studied the closely related problem of the branch orientation
of DLA. We adopt the Horton-Strahler's scheme of branch ordering
\cite{Leopold}.  The smallest branches without side-branch are
assigned order 1 and the main stems have the highest orders. Other
authors have used slightly different schemes \cite{Feder,Ossadnik},
which should give similar results to those reported here. From visual
examinations of the branches shown in Ref. \cite{Feder}, it is evident
that those with the highest order branches are nearly radial and the
directedness decreases for lower orders.  Quantitatively, we define
the branch orientation angle $\phi$ to be measured counter-clockwise
from the position vector of the base of the branch to the branch
orientation vector pointing from the base to the tip.  Figure
\ref{F_branch} plots the probability distribution $P(\phi,s)$ of
$\phi$ for the branch order $s=1$ to 5. The data was averaged over 50
clusters, one million particles each.  The distribution $P(\phi,s)$
can be approximated reasonably well by a cosine curve for $n$ close to
1.  In the cosine expansion of $P(\phi,s)$, the coefficient of the
$cos(\phi)$ term decreases rather quickly with $s$, while that of $cos(2\phi)$
is negative and rising up
towards 0. The coefficients as functions of $s$ are not well described
by simple functional forms.

\section{phenomenological model}

	We now propose an intuitive phenomenological picture for the
properties of the angle of attachment. For convenience in
presentation, we examine the following approach of DLA growth, instead
of the more efficient random walker method. To add a particle to the
cluster, the Laplacian potential is solved to obtain the particle
probability flux lines. The new particle is launched randomly with
uniform probability on a big circle inscribing and far away from the
cluster. It subsequently traces the flux line deterministically
towards the cluster until it hits and become part of it. In this
scheme, the direction of the attachment is simply the tangent to the
flux line at the point of contact. Figure \ref{F_flux} shows 200 flux
lines with uniformly spaced starting positions far away. They are all
equally likely to represent the next growth step. We will concentrate
on the geometry of these typical flux lines which are relevant to the
growth.  Lines with very low probability weight, which can have very
different geometry
\cite{Evertszflux}, are neglected.

	The flux lines proceed nearly exactly radially inwards until
they are close enough to the cluster to be influenced by the geometry
of the main branches.  At this stage, the details of the side-branches
are still unimportant. Depending on the position of the flux line
relative to the nearest main branch, it can keep proceeding radially
towards the tip or make a turn to approach the branch from either
side. After advancing further, the geometry of the nearest
side-branches becomes important.  Again, the flux line approaches
either towards the tip or one of the sides of the side-branch. Similar
situation recurs for side-branches of the side-branches until the
geometry of individual particles becomes relevant.

The typical total number of turns thus equals the number of levels of
side-branching, which is proportional to $\tau=\ln N$
\cite{Feder,Ossadnik}. The angle of attachment is approximately the sum
of the angles of turning. Suppose that when $\tau$ is increased by
$\Delta \tau$, there are, on average, one more turn in the flux lines.
We approximate the corresponding evolution of the distribution of the
angle of attachment by:
\begin{equation}
\label{conv}
P(\theta, \tau + \Delta \tau) = \int_{-\pi}^{\pi} G(\theta-\theta',
\Delta \tau) P(\theta', \tau) d \theta'
\end{equation}
where there is a reparametrization with $\tau$. Implicit assumptions
includes the uncorrelation of the turns and their isotropy so that the
propagator $G$ only depends on $\theta-\theta'$.  Unfortunately, a
direct numerical verification of Eq.  (\ref{conv}) with the
determination of $G$ requires computation of the flux lines for very
large clusters, which is well beyond the capability of the relaxation
method \cite{Viscekbook}. The validity of the assumptions can only be
justified by comparing their predictions with simulations.

We consider first the simplest case where $G$ is a Gaussian centered
at $\Delta\theta=0$ with width smaller than $\pi$. Equation
(\ref{conv}) reduces to:
\begin{equation}
\label{diffusion}
\frac{\partial}{\partial \tau} P(\theta, \tau) =
\nu \frac{\partial^2}{\partial \theta^2} P(\theta, \tau)
\end{equation}
where $\nu$ is the diffusion constant. The set of Eqs (\ref{Pexpan}) and
(\ref{a_n}) is indeed a solution to Eq. (\ref{diffusion}). They predict
$\gamma_n=\nu n^2$ and thus the amplitudes of the higher harmonics
decay much faster. This is consistent with the practically zero values
of the measured amplitudes for $n\ge 3$.  Moreover, the result
$\gamma_2/\gamma_1=4$ is in the same order as the numerical value
6.7(3).  For general $G$, we expand:
\begin{equation}
\label{Gexpan}
G(\Delta \theta) = \frac{1}{2\pi} + \sum_{n=1}^{\infty}
\frac{G_n}{\pi} cos(n\Delta\theta)
\end{equation}
Substitution into Eqs (\ref{Pexpan}) and (\ref{a_n}) gives $\gamma_n=-\ln
G_n/\Delta \tau$. Consider the next simplest form besides the
Gaussian: $G(\Delta \theta) \sim \exp[ - ( \Delta\theta / \theta_0 )
^4]$.  By matching $\gamma_1$ and $\gamma_2$ with the numerical
values, we obtain respectively $\theta_0\equiv 94^\circ$ and $\Delta
\tau=\ln 130$. It means that a flux line make a turn of typically not
more than $94^\circ$ per increase in $N$ by a factor of 130. For a
cluster of $N=10^6$, each flux lines has roughly $N_t\simeq 2.8$
turns. We also tried $G$ in a very different form of the
sum of two Gaussian peaks. It gives $N_t\simeq 4$.

Recall that the number of turns $N_t$ equals approximately the number
of levels of side-branching probed by the flux lines. The maximum
branch order in a cluster of a million particles is typically 9
\cite{Ossadnik}.  However, most flux lines terminate around the tips
of the cluster so that the higher order branches, which reside close
to the center, are irrelevant. As a result, $N_t \ll 9$. It seems
that $N_t\simeq 3$ or 4 obtained from the very crude models are
reasonable approximates.  Even for large clusters, $N_t$ is very
small. This is the cause of the extremely slow evolution of $P(\theta,
N)$.

	The length scale of the problem decreases as we consider
deeper levels of side-branching.  In our construction with the
modified Gaussian, the displacement of the flux line in between turns
typically decreases by a factor of $130^{1/D} \simeq 17$ after every
turn, where $D\simeq 1.72$ is the fractal dimension of DLA
\cite{Viscekbook}. every flux line converges rapidly to its point of
contact. For extremely large clusters and length scales in between
that of the cluster and the individual particles, segments of flux
lines containing their points of contact are statistically similar to
each other after proper rescaling.  This scaling property is a
consequence of the approximation that the propagator $G$ is
independent of $N$, based on the assumption that the lines progress in
self-similar environments.  Dependence of $G$ on $N$ should reveal
in discrepancies from Eq.  (\ref{a_n}), which corresponds to any
non-linearity in the log-log plot in Fig. \ref{F_amp}(a) and (b).
Within our accuracy, we cannot identify any unambiguous non-linearity
for $N\agt 300$. Our measurement is not sensitive to the well known
deviations from self-similarity of DLA
\cite{Viscekbook,Mandelbrot}.

Although the uncorrelation of the turns seems to be a reasonable
approximation, the are obviously strong correlations within a turn so
that Eq. (\ref{conv}) cannot be reformulated with smaller $\Delta
\tau$ corresponding to fractions of a turn. Indeed,
$G(\Delta\theta,\Delta\tau/2)$ is not well defined in both models with
the modified Gaussian and the double Gaussian, since $G_3<0$.

	The behavior of the distribution $P(\phi,s)$ of the branch
orientations can be accounted for similarly. The main branches with
the highest orders are nearly radial. Their side-branches are attached
to either sides with a subtended angle around $\pm 40^\circ$
\cite{Ossadnik}.  This contributes to a step in the randomization of
the orientation analogous to the turns of the flux lines. Therefore,
much of above discussion applies. However, the correlation in the
randomization process is stronger here, since a branch physically
excludes some of the side-branches of smaller orders to have the same
orientation. It is the exclusion by the radial main branches which
causes the dips of $P(\phi,s)$ at $\phi=0$ in Fig.
\ref{F_branch}.  This correlation also lead to the break down of
any analogous power law decay of the amplitudes in the cosine
expansion of $P(\phi,n)$.

\section{Discussions}

	The local anisotropy of DLA is far from being stabilized for
clusters of size $N \alt 10^6$.  Assuming that Eq. (\ref{a_n}) can be
extrapolated to $N\rightarrow \infty$, we get
$\lim_{N\rightarrow\infty}P(\theta,N)=1/2\pi$ and DLA is locally
$isotropic$ asymptotically. The reliability of this extrapolation
deserves special attention due to its simplicity, excellent agreement
with simulation for $300\alt N \alt 10^6$, and consistency with an
intuitive phenomenological model. At finite $N$, the anisotropy can
quantitatively be expressed by the proportion, $P_F$, of forward
sticking events given by $P_F \simeq 0.5+2a_1 N ^ { -
\gamma_1}$ from Eq. (\ref{P}). For $N=10^3$ to $10^6$, $P_F$
decreases from $80\%$ to $65\%$. The anisotropy is still strong and
agrees with the density correlation measurements \cite{Meakin_corr}.
Achieving approximate isotropy with $P_F \alt 55\%$ requires $N \agt
10^{11}$!

	Our results on $P(\theta,N)$ is one of the most accurate
non-trivial measurements ever done on DLA. In particular, the amazing
accuracy in the one parameter fit to the simple analytic form in Eq.
(\ref{P}) at $N=614400$ is rare. The $cos(\theta)$ term, which is the
dominant term representing the anisotropy, was not identified before
because forward and backward attachments are not properly
discriminated in the correlation measurements \cite{Meakin_corr}.

	We hope to motivate similar measurements for diffusion limited
deposition (DLP). If analogous anisotropy tends to zero
asymptotically, very large DLP might not be self-affine
\cite{Meakin_dlp}. The same approach might also be attempted for the
dielectric breakdown model
\cite{Viscekbook}.

	The numerical part of this work was done with Henry Kaufman
and Benoit B. Mandelbrot, who are gratefully acknowledged.

{}~

\begin{figure}
\caption{
Probability distribution $P(\theta,N)$ of the angle of attachment
$\theta$ for different cluster sizes $N$. Also shown are one parameter
cosine fits for the data at $N=9600, 76800$ and 614400 and two parameters
fits for $N=150$ and 1200.
\label{F_Prob}
}
\end{figure}

\begin{figure}
\caption{
Amplitudes $a_n(N)$ of the cosine terms in the expansion of
$P(\theta,N)$ as a function of cluster size $N$: (a) $a_1$ to
$a_4$ in semi-log plot; (b) $a_1$  and (c)  $a_2$ in log-log plots
respectively.
\label{F_amp}
}
\end{figure}

\begin{figure}
\caption{
Probability distribution $P(\phi,s)$ of the branch orientation angle
$\phi$ for branches of order $s$ and fits with four cosine terms.
\label{F_branch}
}
\end{figure}

\begin{figure}
\caption{
Example of 200 particle probability flux lines with uniformly spaced
starting position on a circle of radius $4R_G$, where $R_G$ is radius
of gyration of the one million particles DLA cluster. The potential was
solved by over-relaxation method in a circular region of radius $4R_G$
on a grid with lattice spacing $8R_G/7000$.  Displayed region has
radius $2R_G$.}
\label{F_flux}
\end{figure}


\begin{references}
\bibitem{Witten}
      T. A. Witten and L. M. Sander, Phys. Rev. Lett. {\bf 47},
1400 (1981)., Phys. Rev. B {\bf 27}, 2586 (1983).
\bibitem{Viscekbook}
        T. Vicsek, Fractal growth phenomena, 2nd Ed. (World
Scientific, Singapore 1992).
\bibitem{Kaufman}
H. Kaufman, unpublished.
\bibitem{Meakin_corr}
P. Meakin and T. Viscek, Phys. Rev. A {\bf 32}, 685 (1985).
\bibitem{Mandelbrotbook}
B.B. Mandelbrot, The fractal geometry of nature (Freeman, San
Francisco 1982).
\bibitem{Mandelbrot}
B.B. Mandelbrot and T. Viscek, J. Phys. A {\bf 22}, L377 (1989).
\bibitem{Leopold}
L.B. Leopold, American Scientist {\bf 50}, 511 (1962).
\bibitem{Feder}
J. Feder, E.L. Hinrichsen, K.J. M{\aa}l{\o}y and T. J{\o}ssang,
Physica D {\bf 38}, 104 (1989).
\bibitem{Ossadnik}
P. Ossadnik, Phys. Rev. A {\bf 45}, 1058 (1992).
\bibitem{Evertszflux}
C.J.G. Evertsz and B.B. Mandelbrot, Physica A {\bf 177}, 589 (1991).
\bibitem{Mandelbrot}
B.B. Mandelbrot, Physica A {\bf 191} 95 (191).
\bibitem{Meakin_dlp}
P. Meakin, J. Kert\'{e}sz and T. Vicsek, J. Phys. A{\bf 21} 1271 (1988).
\end{references}
\end{document}